\documentclass[12pt,a4paper]{article}
\usepackage[top=2.3cm,bottom=2.8cm,left=2.1cm,right=2.1cm]{geometry}
\usepackage{verbatim}
\usepackage{latexsym}
\usepackage{amsmath}
\usepackage{amsfonts}
\usepackage{amssymb}
\usepackage{amsthm}
\usepackage{amscd}
\usepackage{mathrsfs}

\newcommand{\be}{\begin{equation}}
\newcommand{\ee}{\end{equation}}
\newcommand{\bea}{\begin{eqnarray}}
\newcommand{\eea}{\end{eqnarray}}

\renewcommand{\theequation}{\arabic{section}.\arabic{equation}}

\def\ad{{\mathrm{ad}}}                  %
\def\Ad{{\mathrm{Ad}}}                  %
\def\H{{\mathcal{H}}}                   %
\def\ri{{\mathrm{i}}}                   %
\def\K{{\mathcal{K}}}                   %
\def\vG{\overrightarrow{G}}             %
\def\vg{\overrightarrow{g}}             %
\def\cG{{\mathcal{G}}}                  %
\def\vcG{\overrightarrow{\cal G}}       %
\def\vX{\overrightarrow{X}}             %
\def\veta{{\overrightarrow{\eta}}}      %
\def\vY{\overrightarrow{Y}}             %
\def\Ker{\mathrm{Ker}}                  %
\def\vlambda{{\overrightarrow{\lambda}}}
\def\vLa{{\overrightarrow{\Lambda}}}    %
\def\vstarLa{{\overrightarrow{\Lambda^*}}}    %
\def\vnu{{\overrightarrow{\nu}}}    %
\def\vq{\overrightarrow{q}}             %
\def\U{{\mathcal{U}}}                   %
\def\cO{{\mathcal{O}}}                  %
\def\vcO{\overrightarrow{\cO}}          %
\def\Act{{\mathrm{C}}}                  %
\def\vJ{\overrightarrow{J}}             %
\def\Actext{\hat{\mathrm{C}}}           %
\def\vp{\overrightarrow{p}}             %
\def\cT{{\mathcal{T}}}                  %
\def\bT{{\mathbb{T}}}                   %
\def\cQ{{\mathcal{Q}}}                  %
\def\bR{{\mathbb{R}}}                   %
\def\vxi{\overrightarrow{\xi}}          %
\def\red{{\mathrm{red}}}                %
\def\diag{{\mathrm{diag}}}              %
\def\cZ{{\mathcal{Z}}}                  %
\newcommand{\calT}{\mathcal{T}}        %
\newcommand{\cU}{\mathcal{U}}          %
\newcommand{\cN}{\mathcal{N}}          %
\newcommand{\calG}{\mathcal{G}}        %
\newcommand{\cP}{\mathcal{P}}          %
\newcommand{\bC}{\mathbb{C}}           %
\newcommand{\bZ}{\mathbb{Z}}           %
\newcommand{\cI}{\mathcal{I}}          %

\begin{document}

\vspace*{0.5cm}
\begin{center}
{\Large \bf  Generalized spin Sutherland systems revisited}
\end{center}

\vspace{0.2cm}

\begin{center}
L. Feh\'er${}^{a,b}$ and B.G. Pusztai${}^{c,d}$  \\

\bigskip

${}^a$ Department of Theoretical Physics, University of Szeged\\
Tisza Lajos krt 84-86, H-6720 Szeged, Hungary \\
${}^b$ Department of Theoretical Physics, Wigner RCP,  RMKI\\
H-1525 Budapest 114, P.O.B. 49,  Hungary\\
e-mail: \texttt{lfeher@physx.u-szeged.hu}

\bigskip

${}^c$ Bolyai Institute, University of Szeged \\
Aradi v\'ertan\'uk tere 1, H-6720 Szeged, Hungary  \\
${}^d$ MTA Lend\"ulet Holographic QFT Group, Wigner RCP \\
H-1525 Budapest 114, P.O.B. 49, Hungary \\
e-mail: \texttt{gpusztai@math.u-szeged.hu}

\bigskip

\end{center}

\vspace{0.2cm}

\begin{abstract}
We present generalizations of the spin Sutherland systems obtained
earlier by Blom and Langmann and by Polychronakos in two different ways:
from $SU(n)$ Yang--Mills theory on the cylinder and by constraining
geodesic motion on the $N$-fold direct product of $SU(n)$ with itself, for any $N>1$.
Our systems are in correspondence with the Dynkin diagram automorphisms of arbitrary
connected and simply connected compact simple
Lie groups.
 We give a finite-dimensional as well as an infinite-dimensional derivation and
  shed light on the mechanism whereby
  they lead
to the same classical integrable systems.
The infinite-dimensional approach, based on
 twisted current algebras (alias Yang--Mills with twisted boundary conditions),
was inspired by the derivation of the spinless Sutherland model due to Gorsky and Nekrasov.
The finite-dimensional method relies on Hamiltonian reduction
under twisted conjugations of $N$-fold direct product groups,
linking the quantum mechanics
of the reduced systems to representation theory similarly as was explored previously
in the $N=1$ case.

\end{abstract}

\newpage

\section{Introduction}
\setcounter{equation}{0}

The  Calogero--Moser--Sutherland type many-body systems \cite{Cal,Sut,Mos} and their generalizations
continuously attract attention since they
are ubiquitous in physical applications and are related to intriguing mathematical structures.
See, for instance, the reviews \cite{RujR,Nekr,PolR,EtiR}.
A powerful approach to these integrable systems consists in viewing them as Hamiltonian reductions
of higher dimensional free systems. This method first came to light in the
papers \cite{OPCim,KKS}. In particular,
Kazhdan, Kostant and Sternberg \cite{KKS} obtained the classical Sutherland Hamiltonian
\be
\H(q,p) = \frac{1}{2} \sum_{k=1}^n p_k^2 +
\sum_{i\neq j} \frac{\nu^2}{\sin^2(q_i-q_j)}
\ee
by reducing the kinetic energy of the geodesic motion
on the group $SU(n)$ using the symmetry defined by the conjugation action of $SU(n)$ on itself.
Their reduction required fixing
the Noether charges of the symmetry in a very special manner.
Later it turned out
that more general choices lead to extensions of the Sutherland system by
`spin' degrees of freedom that belong to reductions of coadjoint orbits of $SU(n)$ by
the action of the maximal toral subgroup.
Spin Sutherland systems of this kind appear for every simple Lie group
\cite{Nekr,LiXu,Res,Hoch}.
Moreover, not only the choice of the constraints, alias the value of the
momentum map,  but also the underlying symmetry  can be
generalized in several ways.
For example, we obtained spin Sutherland systems by reducing the
free motion on simple compact Lie groups utilizing twisted conjugations \cite{FPNucl}.
These reductions were analyzed \cite{FPJPA} at the quantum mechanical level, too, which
yields a bridge from harmonic analysis to integrable systems.

The derivation of the Calogero and Sutherland systems given  in \cite{KKS} represents
a paradigm for the reduction approach to integrable systems. This framework
has since been expanded by the introduction of several new parent
systems as master integrable systems.
We mention only the influential work of Gorsky
and Nekrasov \cite{GN-YM,GN}, where  the Sutherland Hamiltonian was re-derived from a free Hamiltonian
on an infinite-dimensional phase space built on $su(n)$-valued currents on the circle.
As discussed also in \cite{Nekr,La}, this connects the Sutherland system to $SU(n)$ Yang--Mills
theory on the cylinder.

Our present work was motivated mainly by questions stemming from the papers
of Blom and Langmann \cite{BL1,BL2} and Polychronakos \cite{Pol}, where a
family of generalized spin Sutherland systems was derived by means
of two different methods. In these systems the particle positions
are coupled to  spin variables living on $N$ arbitrary coadjoint orbits of $SU(n)$, and
the Hamiltonian also involves $N$ arbitrary scalar parameters, for any integers $n>1$ and $N>1$.
Blom and Langmann obtained their systems from $SU(n)$ Yang--Mills theory on the cylinder,
placing non-dynamical `color charges' at $N$ arbitrary
locations on the circle. Thus their derivation fits in the Gorsky--Nekrasov framework.
On the other hand, Polychronakos proceeded by constraining the geodesic motion on the $N$-fold
direct product of $SU(n)$ with itself, using $N$ arbitrary scale factors in the
definition of the bi-invariant Riemannian metric.
Although it was not stressed in \cite{Pol}, this  amounts  to Hamiltonian
symmetry reduction with respect to twisted conjugations on the
direct product group, defined  with the aid of the cyclic permutation of the $N$-factors.
Based on direct comparison of the Hamiltonians,
both Blom--Langmann \cite{BL2} and Polychronakos \cite{Pol} pointed out that their respective
systems coincide, but they did not provide any conceptual explanation of this remarkable fact.
It is natural to ask for a better understanding of the mechanism behind
this coincidence.

In this  paper we describe two derivations of group theoretic
generalizations of the
 systems of \cite{BL1,Pol} and shed light on the mechanism whereby these
two derivations lead to the same outcome.
The first approach relies on symplectic reduction of
geodesic motion on the $N$-fold direct product $G\times \cdots \times G$ for any
compact simple Lie group $G$.
The underlying symmetry is built from twisted conjugations
involving the cyclic permutation of the $N$ factors and, for simply laced groups,
also a Dynkin diagram
automorphism of $G$.  For $N=1$ these are the reductions analyzed
in \cite{FPNucl,FPJPA}.  The second method utilizes a free Hamiltonian on the
cotangent bundle of an
infinite-dimensional configuration space consisting of Lie algebra valued
(quasi-)periodic currents.
The reduction is based on the natural action of the corresponding
 (twisted) loop group, where the twisting can be non-trivial in the simply laced cases.
This second derivation fits in the framework of Gorsky--Nekrasov \cite{GN-YM,GN,Nekr} and
Blom--Langmann \cite{BL1,BL2}.
In the infinite-dimensional reduction a partial gauge fixing associated with a
finite-dimensional gauge slice will be exhibited that will be explicitly mapped
into a gauge slice of a corresponding partial gauge fixing that
arises in the finite-dimensional reduction.
The mapping between these gauge slices will be seen to be a bijection that
engenders an isomorphism of the respective reduced systems.
When $G=SU(n)$ equipped with the identity diagram automorphism,
then the reduced systems are precisely those found in \cite{BL1,Pol}.

The organization of the paper is as follows.
Section 2 is devoted to group theoretic preliminaries regarding
the relevant twisted conjugations of the direct product groups.
No new results are claimed in this section, even though we could not find
references for the statements of the two technical lemmas
that we prove.
Section 3 contains the analysis of the finite-dimensional
Hamiltonian reduction under twisted conjugations.
The structure of the reduced system is given by
Proposition 3, which represents our first new result.
The formula of the reduced free Hamiltonian displayed by equation (\ref{G23}) is
spelled out in detail in Subsection 3.2 in the case of the
identity diagram automorphism for any group $G$.
Proposition 4 implies that this reproduces the spin Sutherland systems of \cite{BL1,Pol}  for
$G=SU(n)$.
In Section 4 we present an alternative derivation in the general case based on
(twisted) current algebras.
Our second result is that we here explain the equivalence
of the reduced
systems that result from the finite-dimensional and from the infinite-dimensional
derivations.
We conclude in Section 5 by outlining the quantum mechanical analogue
of the finite-dimensional classical Hamiltonian reduction of Section 3.
There is also an appendix, where a useful technical result is described.

\section{Twisted conjugations and the monodromy matrix}
\setcounter{equation}{0}

In this section we collect the group theoretic results that will be needed later.

Let $G$ be a connected and simply-connected compact simple Lie group.
Choose an automorphism $\gamma$ of $G$  that corresponds to a diagram automorphism,
denoted by $\gamma'$,
of the Lie algebra $\cG$ of $G$ with respect to the Cartan subalgebra $\cT < \cG$.
If $\cG$ is not simply laced then $\gamma$ is the identity automorphism,
otherwise it can have order 1, 2 or 3; the latter occurs only for $G=\mathrm{Spin}(8)$.
The group $G$ acts on itself by the twisted conjugations $C_\eta^\gamma$ that for
each $\eta \in G$ operate as
\be
C^\gamma_\eta\colon G \to G,\quad
g \mapsto \gamma(\eta) g \eta^{-1}.
\label{T1}\ee
The space of the corresponding $G$-orbits can be identified \cite{MW} with the space of the orbits of
the so-called twisted Weyl group acting on the fixed point set $\bT^\gamma$ of $\gamma$
in the maximal torus $\bT < G$ associated to $\cT$.
For non-trivial $\gamma$, the $G$-orbits in question are  termed twisted conjugacy classes.
The maximal (also called principal) orbits with respect to the $C^\gamma$ action fill a dense
open subset $G' \subset G$, and  the space of these orbits can be parametrized by the interior of
a convex polytope,
 $\check \cT^\gamma \subset \cT^\gamma$.
This means that every $C^\gamma$ orbit of $G$ in $G'$ passes through a unique element of the form
$e^\mu$ for some $\mu \in \check \cT^\gamma$.  Moreover, the isotropy subgroup of these
elements equals $\bT^\gamma$, that is, if $\mu \in \check \cT^\gamma$ then
$C_\eta^\gamma(e^\mu)= e^{\mu}$ holds if and only if $\eta \in \bT^\gamma$.
The closure of $\check \cT^\gamma$ is often called `Weyl alcove', since
 it is a fundamental domain of the (twisted) affine Weyl
 group  acting on $\cT^\gamma$.
 Its explicit form can be found for example in \cite{Kac} (see also \cite{FPJPA}).

Now pick a positive integer, $N$, and
consider the direct product
\be
\vG:= \overbrace{G \times \cdots \times G}^{\hbox{$N$ times}}.
\ee
Adopting an admittedly unusual (but convenient)  notation,
the elements of $\vG$ will be encoded as $N$-component `vectors'
\be
\vg=g_1 \oplus \cdots \oplus g_N,
\qquad g_k \in G, \quad \forall k=1,\ldots, N.
\ee
The group operation is of course given by componentwise multiplication.
We denote the Lie algebra of $\vG$ as
\be
 \vcG=\overbrace{\cG \oplus \cdots \oplus \cG}^{\hbox{$N$ times}}
\ee
with elements
\be
\vX=X_1 \oplus \cdots \oplus X_N,
\qquad X_k \in \cG, \quad \forall k=1,\ldots, N.
\label{vX}\ee

By combining $\gamma\in \mathrm{Aut}(G)$ with the
cyclic permutation automorphism of $\vG$ we introduce
$\Gamma \in \mathrm{Aut}(\vG)$  by
\be
\Gamma(\eta_1 \oplus \eta_2 \oplus \cdots  \oplus \eta_N) :=
\gamma(\eta_N) \oplus \eta_1\oplus \cdots \oplus \eta_{N-1}
\ee
and let $\Gamma'$ denote the corresponding automorphism of $\vcG$.
We then define an action of
$\vG$ on itself by the $\Gamma$-twisted conjugations
${\Act}^\Gamma_{\veta}$
that for each $\veta\in \vG$ operate on $\vG$ as follows:
\be
{\Act}^\Gamma_{\veta} (\vg) :=
\Gamma(\veta) \vg \veta^{-1} =
\gamma(\eta_N) g_1 \eta_1^{-1} \oplus \eta_1 g_2 \eta_2^{-1}\oplus\cdots
\oplus \eta_{N-1} g_{N} \eta_{N}^{-1}.
\label{2.7}\ee

The $\Gamma$-twisted conjugations on $\vG$ and the $\gamma$-twisted
conjugations on $G$ are related by the  `monodromy map' $M\colon \vG \to G$
given by
\be M(\vg) := g_1  g_{2}
\cdots g_N.
\ee
It is easy to check the equivariance property
\be
M(\Act^\Gamma_{\veta} (\vg) ) = \Act^{\gamma}_{\eta_N}( M(\vg )).
 \ee
This implies immediately that if $\vg$ and $\vg'$ lie on the same orbit with
respect to the $\Act^\Gamma$ action of $\vG$, then their `monodromy
matrices' lie on the same orbit of $G$ under the $\Act^{\gamma}$
action. The next lemma states that the converse is also true.

\medskip\noindent
{\bf Lemma 1.} \emph{The elements $\vg$ and $\vg'$ lie on the same orbit of
$\vG$ under $\Gamma$-twisted conjugations
if and only if $M(\vg)$ and $M(\vg')$ lie on the same $G$-orbit under
$\gamma$-twisted conjugations.
Furthermore, the corresponding isotropy subgroups of $\vg$ in $\vG$ and $M(\vg)$ in $G$ are
isomorphic. }

\begin{proof}
Suppose that $\vg, \vg'\in \vG$ verify
$M(\vg')= \Act^{\gamma}_{\eta_N}( M(\vg ))$ for some $\eta_N\in G$.
After fixing such an element $\eta_N$, the requirement
$\vg'= \Act^\Gamma_{\veta} (\vg)$, which can be spelled out as
\be
g_1'= \gamma(\eta_N) g_1 \eta_1^{-1},\,\,
g_2'= \eta_1 g_2 \eta_2^{-1},
\ldots, g'_{N-1} = \eta_{N-2} g_{N-1} \eta_{N-1}^{-1},\,\,
g'_N= \eta_{N-1} g_N \eta_{N}^{-1},
\ee
permits us to uniquely determine $\eta_1,\ldots, \eta_{N-1}$.
In fact, we simply solve the last $(N-1)$ of these equations for
$\eta_{N-1},\ldots, \eta_1$, and then the first equation is
already guaranteed by $M(\vg')= \Act^{\gamma}_{\eta_N}( M(\vg ))$.
The same calculation for $\vg' = \vg$ shows that
$M(\vg)= \Act^{\gamma}_{\eta_N}( M(\vg ))$ if and only if
$\vg= \Act^\Gamma_{\veta} (\vg)$ with
\be
\eta_{N-1} = g_N \eta_N g_N^{-1},\,
 \eta_{N-2}= (g_{N-1} g_N) \eta_N (g_{N-1} g_N)^{-1},
\ldots,\,\eta_1 = (g_2 \cdots  g_N) \eta_N (g_{2} \cdots  g_N)^{-1},
\ee
which explicitly establishes the isomorphism claimed by the lemma.
\end{proof}

To continue, fix  positive numbers $\lambda_k$ ($k=1,\ldots, N)$ and define
an invariant scalar product on
 $\vcG$ by
\be
\langle \vX, \vY \rangle_\vlambda := \sum_{k=1}^N \lambda_k \langle X_k, Y_k\rangle,
\label{lambda}\ee
where $\langle\ ,\ \rangle$ is a multiple of the Killing form of $\cG$, i.e.,
$\langle X,Y\rangle = - C\operatorname{tr} (\ad_X \circ \ad_Y)$ with a constant $C>0$.
For later convenience, we assume that
\be
\sum_{k=1}^N \frac{1}{\lambda_k}=1.
\label{T13}\ee
Let ${\Gamma'}^T$ be the transpose of $\Gamma'$ with respect to this scalar product.
The defining property
\be
\langle \vX, \Gamma'(\vY) \rangle_\vlambda = \langle {\Gamma'}^T(\vX), \vY \rangle_\vlambda,
\qquad
\forall \vX, \vY \in \vcG,
\ee
entails that
\be
{\Gamma'}^T: X_1\oplus  \cdots \oplus X_{N-1} \oplus
X_N \mapsto \frac{\lambda_2}{\lambda_1} X_2 \oplus\cdots
\oplus \frac{\lambda_N}{\lambda_{N-1}} X_N \oplus
 \frac{\lambda_{1}}{\lambda_N} (\gamma')^{-1}(X_1),
\ee
where we used that $\gamma'$ preserves the Killing form of $\cG$.

We now introduce two Abelian subalgebras $\K$ and $\cQ$ of $\vcG$, which are
isomorphic to $\cT^\gamma$:
\be
\K:= \cT^\gamma_{\mathrm{diag}}= \{ \vX\vert X_1=X_2=\cdots = X_N \in \cT^\gamma\},
\label{T16}\ee
\be
\cQ:= \{ \vq\,\vert\, q_k = \frac{q}{\lambda_k},
\quad q\in \cT^\gamma \quad (\forall k=1,\ldots N) \}.
\label{T17}\ee
We also define
\be
\check \cQ:= \{ \vq\,\vert\, q_k = \frac{q}{\lambda_k},
\quad q\in \check \cT^\gamma \quad (\forall k=1,\ldots N)\}.
\label{T18}\ee
It is straightforward to verify that
\be
\cQ \subseteq \Ker(\Gamma'^T - e^{-\ad_{\vq}})
\qquad\hbox{and}\qquad
{\mathrm{Im}}(\Gamma'^T - e^{-\ad_{\vq}})\subseteq \K^\perp,
\quad
\forall q\in \cT^\gamma,
\label{rels}\ee
where `perp' refers to orthogonal complement with respect to the scalar product (\ref{lambda}).
The following strengthening of these properties will be crucial in our
considerations.

\medskip\noindent
{\bf Lemma 2.} \emph{For any $q\in \check\cT^\gamma$, parametrizing $\vq$ as in
(\ref{T18}), we have
\be
\Ker(\Gamma'^T - e^{-\ad_{\vq}})=\cQ
\quad\hbox{and}\quad
{\mathrm{Im}}(\Gamma'^T - e^{-\ad_{\vq}})=\K^\perp.
\ee
Consequently, the map
\be
\cZ(q):= (\Gamma'^T - e^{-\ad_{\vq}})\vert_{\cQ^\perp}\colon \cQ^\perp \to \K^\perp
\label{T21}\ee
is a linear {\em bijection} for any $q\in \check \cT^\gamma$.}

\begin{proof}
On account of the first relation in (\ref{rels}) and
standard linear algebraic facts,  using also that $\dim(\cQ) = \dim(\K)$,
it is sufficient to prove that
\be
\Ker(\Gamma' - e^{\ad_{\vq}}) = \K,
\quad
\forall q \in \check \cT^\gamma.
\label{kerrel}\ee
Suppose that $\vX\in \vcG$ belongs to this kernel,
which means that
\be
\Gamma'(\vX) = e^{\ad_{\vq}}(\vX).
\label{kereq}\ee
By taking  exponential in the group $\vG$, we conclude that
\be
\Gamma(e^{t \vX}) = e^{t \Gamma'(\vX)} = e^{\vq} e^{t \vX} e^{-\vq},
\quad \forall t \in \bR.
\ee
This in turn means that $e^{t \vX}$ belongs to the $C^\Gamma$ isotropy group of
$e^{\vq}$.

To finish the proof, note that  $M(e^{\vq}) = e^q$ holds for all $q \in \cT^\gamma$.
Therefore we see from (the proof of) Lemma 1 that if $q \in \check \cT^\gamma$, then the isotropy
subgroup of $e^{\vq}$ with respect to the $C^\Gamma$ action is precisely  the connected
subgroup $K$ of $\vG$ associated with the Lie algebra $\K$ (alias the diagonal
embedding of $\bT^\gamma$ into $\vG$).

By combining the above, we have shown that
(\ref{kereq}) with $q\in \check \cT^\gamma$ implies that
$\vX \in \K$, i.e., the relation (\ref{kerrel}) is valid.
\end{proof}

\medskip
\noindent
{\bf Remark.} For $q \in \check \cT^\gamma$, consider the transpose of $\cZ(q)$:
\be
\cZ(q)^T=  (\Gamma' - e^{\ad_{\vq}})\vert_{\K^\perp}\colon \K^\perp \to \cQ^\perp.
\ee
Translated into geometric terms,
the equality ${\mathrm{Im}} (\cZ(q)^T) = \cQ^\perp$ says that
\be
T_{e^{\vq}} C^\Gamma_{\vG}(e^{\vq}) = \cQ^\perp e^{\vq}.
\ee
Here  $C^\Gamma_{\vG}(e^{\vq})$ is the $\vG$-orbit through $e^{\vq}$ and on the right-hand-side we
notationwise `pretend' that we are dealing with a matrix Lie group.
Therefore we have the direct sum decomposition
\be
T_{e^{\vq}} \vG =\vcG e^{\vq}= \cQ e^{\vq} \oplus \cQ^\perp e^{\vq} =
T_{e^{\vq}} Q \oplus T_{e^{\vq}}
C^\Gamma_{\vG}(e^{\vq}),
\qquad
\forall \vq \in \check \cQ,
\ee
where $Q$ is the connected subgroup of $\vG$ with Lie algebra  $\cQ$.
This is an orthogonal decomposition with respect to the bi-invariant Riemannian metric on
$\vG$ induced by the scalar product (\ref{lambda}) on $\vcG$.
It is also worth noting that the map $\cZ(q)$ is equivariant
under the actions of
\be
K=\bT^\gamma_{\mathrm{diag}}
\label{K}\ee
 given by
the corresponding restrictions of the adjoint action of $\vG$ on $\vcG$.

\section{Hamiltonian reduction based on twisted conjugations }
\setcounter{equation}{0}

We are interested in reductions of the free particle moving on the group manifold
$\vG$ equipped with the bi-invariant Riemannian metric
belonging to the scalar product
 (\ref{lambda}) on $\vcG$. This metric is clearly invariant also under
the twisted conjugations ${\Act}^\Gamma_{\veta}$ (\ref{2.7}).
We shall characterize the reductions with respect to this symmetry using the standard
framework of symplectic geometry \cite{OR}.

\subsection{General description for any $\gamma \in \mathrm{Aut}(G)$}

We begin by noting that the phase space of the geodesic motion is the cotangent bundle of $\vG$.
We adopt the identification   $T^* \vG \simeq \vG \times \vcG$ set up
utilizing right-translations  and taking $\vcG$ as the model of $\vcG^*$
by the scalar product $\langle\ ,\ \rangle_\vlambda$.
We also choose a coadjoint orbit of $\vG$,
\be
\vcO:= \cO_1 \oplus \cdots \oplus \cO_N,
\label{G1}\ee
where the $\cO_k$ are coadjoint orbits of $G$.
We then consider the extended phase space
\be
P:= T^* \vG \times \vcO \simeq \vG \times \vcG \times \vcO =\{ (\vg, \vJ, \vxi)\}.
\label{G2}\ee
Like in Section 2, we have
 $\vJ = J_1 \oplus\cdots \oplus J_N$ and $\vxi= \xi_1 \oplus \cdots \oplus \xi_N$.
The phase space $P$ carries the symplectic form
\be
\Omega=
d \langle \vJ, d\vg \vg^{-1}\rangle_{\vlambda}
+\omega \quad\hbox{with}\quad \omega =\sum_{k=1}^N \omega_k,
\label{Omega}\ee
where $\omega_k$ is the natural symplectic form of $\cO_k$.
The orbital part $\omega$ implies the Poisson brackets
\be
\{\langle\vxi,\vX \rangle_{\vlambda} ,\langle\vxi,\vY \rangle_{\vlambda}\} =
\langle\vxi,[\vX,\vY] \rangle_{\vlambda},
\label{G4}\ee
for any constants $\vX, \vY\in \vcG$.
The other non-trivial Poisson brackets between the fundamental variables  can be
displayed as
\be
\{ \vg ,\langle\vJ,\vX \rangle_{\vlambda}\} =
\vX \vg,
\qquad
\{\langle\vJ,\vX \rangle_{\vlambda} ,\langle\vJ,\vY \rangle_{\vlambda}\} =
\langle\vJ,[\vX,\vY] \rangle_{\vlambda}.
\label{G5}\ee
Here,  we have  $\vX \vg = X_1 g_1 \oplus \cdots \oplus X_N g_N$.
The Hamiltonian $H$ of the free particle, trivially extended to $P$,
reads
\be
H(\vg, \vJ, \vxi) = \frac{1}{2} \langle \vJ, \vJ \rangle_{\vlambda}.
\label{G6}\ee
The free flow generated by $H$ through the initial value $(\vg, \vJ, \vxi)$ is
\be
(\vg(t), \vJ(t), \vxi(t)) = (e^{t \vJ} \vg, \vJ, \vxi ).
\label{G7}\ee
The twisted conjugations $C_{\veta}^\Gamma$ lift to the
 Hamiltonian action  $\Actext^\Gamma$ of $\vG$
on $P$:
\be
\Actext^\Gamma_{\veta}(\vg,\vJ,\vxi)=
(\Act^\Gamma_{\veta} (\vg), \Ad_{\Gamma(\veta)}(\vJ), \Ad_\veta(\vxi)).
\label{extact}\ee
This action is generated by the equivariant momentum map $\Psi\colon P \to \cG$
given by
\be
\Psi(\vg,\vJ, \vxi) = \Gamma'^T(\vJ) - \Ad_{\vg^{-1}} (\vJ) + \vxi.
\ee
Indeed, the Hamiltonian vector field of the function
$\Psi_{\vX} =\langle\Psi, \vX  \rangle_{\vlambda}$ coincides
with the infinitesimal generator of the $\vG$-action (\ref{extact}).

According to the standard shifting trick \cite{OR},
we can represent any
reduction of the free motion  with respect to the twisted conjugation symmetry
as symplectic reduction of the
Hamiltonian system $(P, \Omega, H)$
at the zero value of the momentum map; in association with some  $\vcO$.
The resulting reduced Hamiltonian system
 lives on the space of $\vG$-orbits
\be
P_{\red}= P_{\Psi=0}/ \vG,
\label{G10}\ee
which inherits a reduced symplectic structure, $\Omega_\red$, and Hamiltonian, $H_\red$,
from $\Omega$ and $H$ on $P$.
To be more precise, one should note that $P_{\red}$ is not always a smooth manifold,
but is in general a so-called stratified symplectic space, i.e., a disjoint union of symplectic
manifolds of various dimensions \cite{OR}.  The detailed structure depends on the choice of the orbit
$\vcO$. We here will not dwell on this issue, but shall instead focus on a dense open
subset $P'$ of the phase space $P$ whose reduction can be characterized rather directly.

Remember that every element of $\vG$ can be transformed by twisted conjugation
into a unique element of the form $e^{\vq}$, parametrized according to (\ref{T18})
by $q$ from the closure
of the generalized Weyl alcove $\check \cT^\gamma$.  From now on we restrict our attention
to the subset $P'$ consisting of triples $(\vg, \vJ, \vxi)$, where the twisted conjugates
of $\vg$ can be parametrized by elements $q$ of the open alcove $\check \cT^\gamma$.
It is then clear that every $\vG$-orbit in $P'_{\Psi=0}$ intersects the subset $S\subset P$
defined by
\be
S:= \{ (e^{\vq}, \vJ, \vxi) \mid \Psi(e^{\vq}, \vJ, \vxi)=0,\quad q \in \check \cT^\gamma\,\}.
\label{G11}\ee
Observe also from Lemma 1 that any $\vG$-orbit in $P'_{\Psi=0}$ intersects $S$
in an orbit of the subgroup $K < \vG$ displayed in (\ref{K}).
 In other words, $S$ is the `gauge slice'
of a partial gauge fixing  of the $\vG$-action on $P'_{\Psi=0}$, for which the
`residual gauge transformations' belong precisely to the subgroup $K$.
In this way we obtain the identification
\be
P_{\red}':= P'_{\Psi=0}/ \vG \simeq S/K.
\label{G12}\ee
Consequently, the $\vG$-invariant smooth functions on $P'_{\Psi=0}$, which descend to
smooth functions on $P'_\red$, are equivalent to the $K$-invariant smooth functions on $S$.
The Poisson brackets of these functions can be determined with the aid of the
restriction of the symplectic form $\Omega$ to $S$.

In order to make the above concrete, we proceed to solve the restriction of the
momentum map constraint on $S$. This amounts to the following requirement:
\be
(\Gamma'^T- e^{-\ad_{\vq}})(\vJ) + \vxi=0.
\label{G13}\ee
To solve this equation we use the decompositions
\be
\K + \K^\perp = \vcG = \cQ + \cQ^\perp
\ee
and write accordingly
\be
\vxi= \vxi_\K + \vxi_{\K^\perp}
\quad\hbox{and}\quad
\vJ = \vJ_\cQ + \vJ_{\cQ^\perp}.
\label{G15}\ee
Here, recall the definitions of $\K$ and $\cQ$ from equations (\ref{T16}) and (\ref{T17}).
Referring to Lemma 2,  the $\K$-component of the
momentum map constraint (\ref{G13}) enforces that
$\vxi_\K=0$, while its $\K^\perp$-component
is completely solved by
\be
\vJ_{\cQ^\perp}= - \cZ(q)^{-1} ( \vxi_{\K^\perp})
\ee
using the operator $\cZ(q)$ defined in (\ref{T21}).
The element $q\in \check \cT^\gamma$ can be chosen arbitrarily, and
$\vJ_\cQ$ is also a free variable, which we parametrize as
\be
\vJ_\cQ= \vp= \frac{p}{\lambda_1} \oplus \frac{p}{\lambda_2} \oplus \cdots \oplus
\frac{p}{\lambda_N},
\qquad
 p \in \cT^\gamma.
\label{G17}\ee
Since the variables $q, p$ and $\vxi_{\K^\perp}$ uniquely label the points of $S$,
we obtain the identification of manifolds
\be
S \simeq \check \cT^\gamma \times \cT^\gamma \times (\vcO\cap \K^\perp) =
\{ (q,p,\vxi_{\K^\perp})\}.
\label{G18}\ee
According to the general principles of Hamiltonian reduction, the
reduced system on $P'_\red$ is determined by the pull-backs of the symplectic form
$\Omega$ and the Hamiltonian $H$ to $S$. Let us denote these pull-backs by $\Omega_{S}$ and
$H_S$. In terms of the model (\ref{G18}) we obtain
\be
\Omega_S = d \langle p, dq \rangle + \omega_{\vcO \cap \K^\perp}.
\ee
Here $\omega_{\vcO \cap \K^\perp}$ is the contribution of the orbital form
$\omega$ from (\ref{Omega}).
The first term is the natural symplectic form of the cotangent bundle
\be
T^* \check \cT^\gamma \simeq \check \cT^\gamma \times \cT^\gamma = \{ (q,p)\}.
\ee
Since the residual gauge transformations by $K= \bT^\gamma_{\diag}$ act only on
$\vxi_{\K^\perp}$,  the reduced phase space can be decomposed according to
\be
P'_{\red} \simeq T^* \check \cT^\gamma  \times (\vcO \cap \K^\perp)/K.
\label{G21}\ee
This is a symplectic identification in the following sense.
The natural action of the group $K < \vG$ on $\vcO$  is generated by the momentum map
$\vxi \mapsto \vxi_\K$, and
\be
\vcO_\red = (\vcO \cap \K^\perp)/K
\label{G22}\ee
is the reduced orbit obtained by setting this momentum map to zero.
We denote the (in general stratified) symplectic structure inherited by
$\vcO_\red$   as $\omega_\red$. This is carried by the second factor in the Cartesian
product (\ref{G21}).
Regarding the reduced Hamiltonian of the free motion, we already mentioned that it is
determined by $H_S$.   By means of the model (\ref{G18}), $H_S$ takes the form
\be
H_S = \frac{1}{2} \langle p, p \rangle
+ \frac{1}{2}
\langle \cZ(q)^{-1} (\vxi_{\K^\perp}), \cZ(q)^{-1} (\vxi_{\K^\perp})\rangle_{\vlambda}.
\label{G23}\ee
The first term is the kinetic energy of the free motion on $\check \cT^\gamma$ with
respect to the flat metric,  while the second (also $K$-invariant) term describes the interaction
of the spatial degrees of freedom $q, p$  with the orbital variables given by the
$K$-orbits in (\ref{G22}).
The outcome of the foregoing discussion can be summarized as follows.

\medskip\noindent
{\bf Proposition 3.} \emph{A model of the open subset of the reduced phase space (\ref{G10})
associated with generic monodromy matrices $M(\vg)$, parametrized by $q\in \check \cT^\gamma$,
is provided by  $T^*\check \cT^\gamma \times \vcO_\red$. Here
$T^* \check \cT^\gamma$ carries its Darboux form and
$\vcO_\red$ (\ref{G22}) is equipped with its own (stratified) symplectic structure
arising by reduction via the $K$-action.
The reduced free Hamiltonian on this subset $P'_\red=S/K$ is encoded by the $K$-invariant
function $H_S$  (\ref{G23}).}

\medskip

We finish this subsection with a few comments.
First, we remark that the Poisson bracket, $\{\ ,\ \}_\red$, that is induced
on the smooth functions on $P'_\red$ is encoded by
\be
\{\langle q, X \rangle, \langle p, Y\rangle \}_\red = \langle X, Y \rangle,
\qquad
\forall X, Y \in \cT^\gamma,
\label{G24}\ee
together with the restriction of the Poisson brackets of the $K$-invariant functions on $\vcO$
to $\vcO \cap \K^\perp$, which can be determined by means of the Lie--Poisson brackets (\ref{G4}).

Second, we recall that the $N=1$ special case of our classical Hamiltonian
reduction was previously studied in \cite{FPNucl}
and the corresponding quantum Hamiltonian reduction was investigated in \cite{FPJPA}.
As will be discussed in Section 5,
the quantum Hamiltonian reduction can be described similarly in the general case.
To facilitate the comparison with these earlier works, we now recast the inverse of
 $\cZ(q)$ (\ref{T21}) in the form
\be
\cZ(q)^{-1} = - e^{\ad_{\vq}} \circ (\U(q)^T)^{-1},
\label{G25}\ee
 where the linear operator $\U(q): \K^\perp \to \cQ^\perp$ reads
\be
\label{G26}\U(q) =
\left(\mathrm{\operatorname{id}}_{\vcG} -
 e^{- \ad_{\vq}}\circ \Gamma'\right)\vert_{\K^\perp}.
\ee
Since $e^{\ad_{\vq}}$ preserves the scalar product on $\vcG$, and inversion and
taking transpose
commute, we can rewrite the Hamiltonian $H_S$ (\ref{G23}) as
\be
H_S = \frac{1}{2} \langle p, p \rangle
+ \frac{1}{2}
\langle \vxi_{\K^\perp}, (\U(q)^T \U(q))^{-1} (\vxi_{\K^\perp})\rangle_{\vlambda}.
\label{G27}\ee
For $N=1$, when $\cQ = \K$,
it was shown in \cite{FPNucl} that the extension of $\U(q)^{-1}$ by zero on $\cQ$ is a solution
of the modified classical dynamical Yang--Baxter equation.
There it was important that the automorphism $\gamma'$ preserves the scalar
product on the underlying Lie algebra. This holds for our $\Gamma'$
if and only if $\lambda_1 = \lambda_2 = \ldots =\lambda_N$.
It would be interesting to know if $\U(q)$ is still related to a classical dynamical
$r$-matrix for any $N$ with arbitrary parameters $\vlambda$.

Third, it is proper to mention  here that Hochgerner \cite{Hoch} studied reductions of cotangent
bundles under the assumption that a single isotropy type occurs for the
underlying group action on the configuration space.
Our Proposition 3 can be obtained as a special case of his results.
Since this would necessitate going into several technicalities,
we have preferred to give a simple  direct derivation.

Finally, note that the reduced system governed by the  Hamiltonian
(\ref{G27}) is exactly solvable since its solutions can be obtained by the projection method
applied to the free flow (\ref{G7}), which involves only algebraic operations.
There exist general arguments \cite{Nekr,Res,Zu} that also indicate Liouville integrability on
the full reduced phase space $P_\red$.
Indeed, one can easily construct many conserved quantities in involution as follows.
For each fixed real parameter $u\neq 0$, define the $\vcG$-valued function $\phi_u$ on the
unreduced phase space by
\be
\phi_u(\vg, \vJ, \vxi):= - \vg^{-1} \vJ \vg + \frac{\vxi}{u}.
\label{G28}\ee
Taking a  basis $\{V^a\}$ of $\vcG$, the Poisson brackets of the components
$\phi^a_u:= \langle \phi_u, V^a\rangle_{\vlambda}$ satisfy
\be
\{ \phi^a_u, \phi^b_v\} = f^{ab}_c\left(\frac{u-1}{u-v} \phi_u^c + \frac{v-1}{v-u} \phi_v^c\right)
\label{G29}\ee
for any $u\neq v$, where $[V^a, V^b] = f^{ab}_c V^c$ and summation over the index $c$ is understood.
It immediately follows from (\ref{G29}) that for any two $\vG$-invariant smooth functions $h_1$ and $h_2$ on $\vcG$
and for any  parameters $u$ and $v$ we have
\be
\{ h_1 \circ \phi_u, h_2 \circ \phi_v\} =0.
\label{G30}\ee
This  descends to a Poisson commuting family on the reduced phase space.
As generators, it is useful to take invariant homogeneous
polynomials $h_i$ and extract the coefficients of the powers of $u^{-1}$  from $h_i\circ \phi_u$.
Because (\ref{G30}) holds for any $u$ and $v$, the so obtained coefficient functions are also in involution.
The $u$-independent term coming from $h(\vX) = \frac{1}{2} \langle \vX, \vX\rangle_{\vlambda}$
gives the main Hamiltonian after reduction, and
it is natural to expect that the family just delineated is generically  sufficient to ensure its
Liouville integrability.

\subsection{The examples associated with $\gamma = \operatorname{id}_G$}

Our goal now is to provide
an explicit formula for the reduced Hamiltonian $H_S$ (\ref{G27}) in
the case of the identity automorphism $\gamma$ of
an arbitrary connected and simply-connected compact
simple Lie group $G$.
During the calculations it proves to be handy to realize
the real Lie algebra $\calG$ as a compact real form of its
complexification $\calG^\bC$. Since the complexification $\calT^\bC$ of
$\calT$ is a Cartan subalgebra of $\calG^\bC$, the pair
$(\calG^\bC, \calT^\bC)$ uniquely determines a reduced root system $\Phi$.
Choose a polarization $\Phi = \Phi_+ \cup \Phi_-$ and let
$\Pi \subset \Phi_+$ denote the corresponding set of simple roots,
whose number is the rank of $G$, $r$.
Also,
for each root $\varphi$ we select an appropriate root vector $X_\varphi$. On
the root vectors we may and shall
impose the following two conditions:
\begin{itemize}

    \item[(a)]
        The root vectors are normalized by the conditions
        \be
            \kappa(X_\varphi, X_{-\varphi}) = 1 \qquad (\varphi \in \Phi_+),
        \label{root_vector_normalization}
        \ee
        where $\kappa$ denotes a positive multiple of the Killing form of $\calG^\bC$.

    \item[(b)]
        The real Lie algebra $\calG$  decomposes as
        \be
            \calG
            = \calT
                \oplus
                \left( \oplus_{\varphi \in \Phi_+} \bR Y_\varphi \right)
                \oplus
                \left( \oplus_{\varphi \in \Phi_+} \bR Z_\varphi \right),
        \label{cG_decomposition}
        \ee
        where for each $\varphi \in \Phi_+$ we define
        \be
            Y_\varphi = \ri \frac{X_\varphi + X_{-\varphi}}{\sqrt{2}}
            \quad \text{and} \quad
            Z_\varphi = \frac{X_\varphi - X_{-\varphi}}{\sqrt{2}}.
        \label{Y_and_Z}
        \ee

\end{itemize}
Take an arbitrary set of vectors $\{ T_j \}_{j = 1}^r \subset \calT$
satisfying $\kappa(T_j, T_k) = -\delta_{j, k}$ $(1 \leq j, k \leq r)$. Clearly
the family of vectors $T_j$  $(1 \leq j \leq r)$ and $Y_\varphi$,
$Z_\varphi$ $(\varphi \in \Phi_+)$ gives an orthonormal basis in $\calG$
with respect to the $\Ad$-invariant Euclidean scalar product obtained by restricting the complex bilinear form
$\langle \ ,\ \rangle = -\kappa$ onto $\calG$.
Obviously, this convention agrees with  the one used in  (\ref{lambda}) and
the definition of
 $\langle\ ,\ \rangle_{\vlambda}$ extends by linearity to the 
 complexification of $\vcG$.

Recalling the form of the Hamiltonian $H_S$ (\ref{G27}), first we wish to
examine the action of the linear operator $\cU(q)^T \cU(q)$
($q \in \check{\calT}$, an open Weyl alcove) on a convenient basis of $\mathcal{K}^\perp$.
For this reason, for all $V \in \calG^\bC$ and $1 \leq I \leq N$
we define the vector $\overrightarrow{V}^I \in \vcG^\bC$ with components
$(\overrightarrow{V}^I)_J = \delta_J^I V$ $(1 \leq J \leq N)$.
To clarify our notations, we remark that for
 $\vX$ (\ref{vX}) we have $(\vX)_I = X_I$.
The family of vectors
\be
     \overrightarrow{T_j}^I \quad (1 \leq j \leq r, \, 1 \leq I \leq N)
    \quad \text{and} \quad
    \overrightarrow{X_\varphi}^I \quad (\varphi \in \Phi, \, 1 \leq I \leq N)
\label{ext_basis}
\ee
gives a basis in the complex Lie algebra $\vcG^\bC$. Now we record that for
each $q \in \calT$  the action of the complexification
of the linear operator
\be
    \cI(q): =
    \left(
        \text{id}_{\overrightarrow{\calG}}
            - e^{- \ad_{\overrightarrow{q}}} \circ \Gamma'
    \right )^T
    \circ
    \left(
        \text{id}_{\overrightarrow{\calG}}
            - e^{- \ad_{\overrightarrow{  q}}} \circ \Gamma'
    \right )
    \colon \vcG \rightarrow \vcG
\label{cI}
\ee
on the  basis (\ref{ext_basis}) reads
\be
    \cI(q) \overrightarrow{T_j}^J
    = \sum_{I = 1}^N
        \frac{1}{\lambda_I}
        M_{I, J}(0)
        \overrightarrow{T_j}^I
    \quad \text{and} \quad
    \cI(q) \overrightarrow{X_\varphi}^J
    = \sum_{I = 1}^N
        \frac{1}{\lambda_I}
        M_{I, J}(\varphi(q))
        \overrightarrow{X_\varphi}^I,
\label{cI_on_basis}
\ee
where $M(\ri x)$ is the $N \times N$ complex matrix with entries
\be
    M_{I, J}(\ri x) =
    \left(
        \lambda_I + \lambda_{I+1}
    \right) \delta_{I, J}
    - \lambda_I e^{- \ri x / \lambda_I} \delta_{I - 1, J}
    - \lambda_{I + 1} e^{\ri x / \lambda_{I + 1}} \delta_{I + 1, J}
    \qquad
    (1 \leq I, J \leq N),
\label{M_entries}
\ee
depending on the real parameter $x \in \bR$.
Here and below,
the capital indices are understood modulo $N$.
The factor $\ri$ occurs since $\varphi (q) \in \ri \bR$ for
each $q \in \calT$ and $\varphi \in \Phi$.
 It is also worth mentioning that the function
$\bR \ni x \mapsto M(\ri x) \in \bC^{N \times N}$ is $2 \pi$-periodic with
symmetry properties
\be
    M_{J, I}(\ri x) = M_{I, J}(-\ri x) = \overline{M_{I, J}(\ri x)}.
\label{M_symmetries}
\ee

On account of the formula (\ref{G27}) and the relationship
\be
    (\cU( q)^T \cU( q))^{-1}
    = (\cI(q)|_{\mathcal{K}^\perp})^{-1},
\label{U_and_I}
\ee
we need  control over the inverse of $M(\ri x)$. One can verify
that $M(\ri x)$ is invertible if and only if $x \in \bR \setminus 2 \pi \bZ$.
Moreover, upon setting
\be
    b_I = \sum_{k = 1}^I \frac{1}{\lambda_k}
    \quad \text{and} \quad
    b_{I, J} = b_I - b_J
    \qquad
    (1 \leq I, J \leq N),
\label{b_def}
\ee
for all $x \in \bR \setminus 2 \pi \bZ$ the entries of the inverse
matrix $\cP(\ri x) = M(\ri x)^{-1}$ can be checked to be
\be
    \cP_{I, J}(\ri x) =
    e^{-\ri b_{I, J} x}
    \left(
        \frac{1}{4} \frac{1}{\sin^2 \left( x / 2 \right)}
        + \ri \frac{b_{I, J}}{2} \cot \left( x / 2 \right)
        -\frac{\vert b_{I, J} \vert}{2}
    \right)
    \qquad
    (1 \leq I, J \leq N).
\label{cP_entries}
\ee
This matrix inversion was treated in detail in \cite{Pol}.
We note that $\varphi(q) \in \ri (\bR \setminus 2\pi \bZ)$ holds
for every $q$ from the open Weyl alcove $\check \cT$ and $\varphi \in \Phi$.

Turning to the study of the degenerate matrix $M(0)$, let us endow the
linear space of the complex column vectors $\bC^N$ with the inner product
\be
    \bC^N \times \bC^N \ni (\boldsymbol{x}, \boldsymbol{y})
        \mapsto
        \boldsymbol{x}^* \Lambda \boldsymbol{y} \in \bC
\label{inner_prod_on_n-space}
\ee
induced by the positive definite diagonal matrix
\be
    \Lambda = \diag(\lambda_1, \ldots, \lambda_N).
\label{Lambda}
\ee
Thinking of the $N \times N$ complex matrices as linear operators acting
from the left on the inner product space $\bC^N$, from the formulae
(\ref{M_entries}) and (\ref{cP_entries}) we can easily deduce the following:
\begin{itemize}

    \item[(i)]
        The operator $\Lambda^{-1} M(0)$ is self-adjoint with respect to the
        inner product (\ref{inner_prod_on_n-space}), the kernel
        $\cN = \ker(\Lambda^{-1} M(0))$ is spanned by the column vector
        $\boldsymbol{n} = [ 1, \ldots, 1 ]^*$, and the orthogonal complement
        $\cN^\perp$ of $\cN$ is an invariant subspace.

    \item[(ii)]
        The restriction
        $(\Lambda^{-1} M(0))|_{\cN^\perp}
            \colon \cN^\perp \rightarrow \cN^\perp$
        is invertible and for its inverse we have
        \be
            \left( (\Lambda^{-1} M(0))|_{\cN^\perp} \right)^{-1}
            = (\pi_{\perp} \cP' \Lambda \pi_{\perp})|_{\cN^\perp},
        \label{on_M0_inverse}
        \ee
        where the $N \times N$ matrix $\pi_{\perp}$ denotes the orthogonal
        projection onto the subspace $\cN^\perp$, whereas $\cP'$ is the
        $N \times N$ matrix with entries
        \be
            \cP'_{I, J} = \frac{(b_{I, J})^2 - \vert b_{I, J} \vert}{2}
            \qquad
            (1 \leq I, J \leq N).
        \label{cP'_entries}
        \ee

\end{itemize}
The operator defined by (\ref{on_M0_inverse}) encodes the action
of the inverse (\ref{U_and_I}) on $\K^\perp \cap \overrightarrow{\cT}$.
It is reassuring to see that the outcome of the elementary algebraic
manipulations is consistent with the statements in (\ref{rels}) and Lemma 2.

Finally, let us observe that due to the relations (\ref{U_and_I}) and
(\ref{cI_on_basis}) we now fully control the linear
operator $(\cU(q)^T \cU(q))^{-1}$.
This implies the explicit formula of $H_S$ given below.

\medskip\noindent
\textbf{Proposition 4.} \emph{For all $N \geq 1$, at each point
$(q, p, \overrightarrow{\xi}_{\mathcal{K}^\perp})
\in \check{\calT} \times \calT \times \overrightarrow{\mathcal{O}}_{\mathrm{red}}$,
the reduced Hamiltonian $H_S$ (\ref{G27}) associated with the trivial automorphism
$\gamma = \text{id}_G$ takes the form
\be
\begin{split}
    H_S %
    =
     \frac{1}{2} \langle p, p \rangle
    + \frac{1}{2} \sum_{j = 1}^r \sum_{I, J = 1}^N
        \cP'_{I, J}
        \langle \overrightarrow{T_j}^I,
            \overrightarrow{\xi}_{\mathcal{K}^\perp}
        \rangle_{\overrightarrow{\lambda}}
        \langle \overrightarrow{T_j}^J,
            \overrightarrow{\xi}_{\mathcal{K}^\perp}
        \rangle_{\overrightarrow{\lambda}}
    \\
    - \frac{1}{2} \sum_{\varphi \in \Phi} \sum_{I, J = 1}^N
        \cP_{I, J}(\varphi(q))
        \langle \overrightarrow{X_\varphi}^I,
            \overrightarrow{\xi}_{\mathcal{K}^\perp}
        \rangle_{\overrightarrow{\lambda}}
        \langle \overrightarrow{X_{-\varphi}}^J,
            \overrightarrow{\xi}_{\mathcal{K}^\perp}
        \rangle_{\overrightarrow{\lambda}}
\end{split}
\label{HSfin}\ee
with $\cP$ and $\cP'$ defined in (\ref{cP_entries}) and
(\ref{cP'_entries}), respectively.
}
\medskip

The above Hamiltonian $H_S$ can be interpreted in terms of `particles'
with positions defined by the components of $q$
that interact with each other as well as with the `spin variables' represented by
$\vxi$. It is a system of  `Sutherland type' since the
interaction exhibits trigonometric dependence on the coordinates, which
means that the `particles' move on the circle.
For $\cG = su(r+1)$, the   explicit formula (\ref{HSfin}) reproduces
precisely the integrable spin Sutherland Hamiltonian obtained previously
by Blom and Langmann \cite{BL1,BL2} and
by Polychronakos \cite{Pol} by means of different methods,
and generalizes their system for arbitrary simple Lie algebras.

\section{Alternative derivation from twisted current algebra}
\setcounter{equation}{0}

It is remarkable that certain many-body systems can be
obtained by alternative reduction procedures starting from finite-dimensional
as well as from infinite-dimensional `free systems'.
The main ideas in this regard go back to
Gorsky and Nekrasov \cite{GN-YM,GN}, who first interpreted
the standard trigonometric Sutherland system from an infinite-dimensional
standpoint.
Their derivation had been exposed in \cite{Nekr,Khes}, too.
In effect, it also formed the
basis of the derivation of the $\cG=su(r+1)$ special case
 of $H_S$ (\ref{HSfin})
given by Blom and Langmann \cite{BL1,BL2}.
We thought it worthwhile to develop the Gorsky--Nekrasov method
in our general case. Below we re-derive the same systems as those
obtained in Section 3.1,  but
we will not be fully rigorous as we shall neglect topological
details concerning the infinite-dimensional spaces involved.

We start with the infinite-dimensional phase space $\widetilde P$ consisting of
triples of $\cG$-valued functions $(A(x), E(x), \zeta(x))$
on the real line, subject to the following
quasi-periodicity condition for $C=A,E, \zeta$:
\be
C(x+1) = \tau'(C(x))
\quad\hbox{with}\quad
\tau:= \gamma^{-1}.
\ee
We assume that $\zeta(x)$ is actually a generalized function of the form
\be
\zeta(x) = \sum_{k\in \bZ} \zeta_k \delta(x-x_k),
\ee
where
\be
0< x_1 < x_2 < \cdots < x_N < 1
\ee
and
\be
x_{k+N} = x_k + 1,
\quad
\zeta_{k+N} = \tau'(\zeta_k), \quad \forall k\in \bZ.
\ee
The normalization is such that  $\int_0^1 \delta(x-x_k) dx =1$ for $k=1,\ldots, N$.
Moreover, we assume that $\zeta_k$ belongs to an adjoint orbit $\cO_k'$ of $G$ in $\cG$,
identified with $\cG^*$ via the scalar product $\langle\ ,\ \rangle$.
Regarding the quasi-periodic $\cG$-valued functions $A$ and $E$, we suppose that they
are smooth on each open interval $(x_k, x_{k+1})$, and they as well as  their
derivatives of all orders have finite one-sided limits at every point $x_k$. For reasons
that will become clear shortly, we permit both $A$, $E$ and all their derivatives
to have jumps at the distinguished points $x_k$.

By using smooth $\cG$-valued quasi-periodic test functions $V(x)$ and $W(x)$,
the non-vanishing  Poisson brackets on $\widetilde P$ are specified as
\be
\Bigl\{ \int_0^1 \langle A(x), V(x)\rangle dx, \int_0^1 \langle E(x), W(x)\rangle dx \Bigr\}
= \int_0^1 \langle V(x), W(x) \rangle dx
\label{4.5}\ee
and
\be
\{ \langle \zeta_k,  V_k \rangle , \langle \zeta_k, W_k \rangle  \} =
\langle \zeta_k , [V_k, W_k] \rangle,
\label{4.6}\ee
where $V_k:= V(x_k)$, $W_k:= W(x_k)$.
The corresponding symplectic form  $\widetilde \Omega$ on $\widetilde P$
reads
\be
\widetilde \Omega = d \widetilde \theta + \omega'
\quad\hbox{with}\quad
\omega' = \sum_{k=1}^N \omega_k',
\label{tildeOm}\ee
where $\omega'_k$ is the Kirillov--Kostant--Souriau form on the orbit $\cO'_k$ and
\be
\widetilde \theta=\int_{0}^1 \langle E(x), d A(x) \rangle dx.
\label{tildetheta}\ee
We equip the phase space $\widetilde P$ with the free Hamiltonian
\be
\widetilde H:= \frac{1}{2} \int_0^1 \langle E(x), E(x)\rangle dx
\label{tildeHam}\ee
and a Hamiltonian action of a suitable twisted loop group, $\widetilde G$.
Namely, an element $g \in \widetilde G$, which is a $G$-valued
function
on the line subject to $g(x+1) = \tau(g(x))$, acts on the triple $(A,E,\zeta)$ according to
\be
A(x) \mapsto g(x) A(x) g^{-1}(x) -  g'(x) g^{-1}(x),
\,\,
E(x) \mapsto g(x) E(x) g^{-1}(x),
\,\,
\zeta_k \mapsto g(x_k) \zeta_k g^{-1}(x_k).
\label{tildeact}\ee
We suppose that $g$ and all its derivatives are smooth on each open interval $(x_k,x_{k+1})$
and have finite one-sided limits similarly to $A$ and $E$. An important difference is that
$g$ is required to be globally continuous on $\bR$ (including at the points $x_k$).
These assumptions guarantee that the formulae in   (\ref{tildeact}) yield a well-defined
action of $\widetilde{G}$ on $\widetilde{P}$.

The action (\ref{tildeact}) admits the equivariant momentum map
\be
\widetilde \Psi(x):=  [A(x), E(x) ] + E'(x) + \zeta(x)
\ee
that takes its values in the distributional dual of the
space of $\cG$-valued smooth quasi-periodic functions.
In particular, the derivative of $E(x)$ is
understood in the distribution sense, i.e., the value $\widetilde \Psi[V]$ of the
functional $\widetilde \Psi$ on
the test function $V$ is
\be
\widetilde \Psi[V]=
\int_{0}^1
\left(\langle V(x), [A(x), E(x)] \rangle  - \langle V'(x), E(x)\rangle\right)dx
+\sum_{k=1}^N \langle V(x_k), \zeta_k \rangle .
\ee
By using this,
one can  verify that $\widetilde \Psi$
generates the infinitesimal action according to
\bea
&&\{ A(x), \widetilde \Psi[V] \} = [V(x), A(x) ] - V'(x),\nonumber\\
&&\{ E(x), \widetilde \Psi[V]\} = [ V(x), E(x)],\\
&&\{ \zeta_k, \widetilde \Psi[V] \} = [V(x_k), \zeta_k].\nonumber
\eea
The reduction of our concern is defined by imposing the momentum map constraint
\be
\widetilde \Psi = 0
\ee
and then factorizing by $\widetilde G$, as usual.
An important fact used in the subsequent analysis is that for every $A$ that appears
in $\widetilde{P}$ there exists an element $g\in \widetilde{G}$ that transforms it
into a constant function on $\bR$.  The constant can be taken from the closure
of the open `twisted Weyl alcove' $\check \cT^\gamma$ that parametrizes the
twisted conjugacy classes in $G$ given by the orbits of the $C^\gamma$-action (\ref{T1}).
For convenience, we present a proof of this fact in Appendix A.

Similarly as we worked in Section 3,  we restrict attention to the subset $\widetilde P'$
consisting of such triples for which $A(x)$ can be gauge transformed to a constant from
the interior of the alcove, denoted as
\be
\chi \in \check \cT^\gamma.
\ee
Mimicking  Section 3.1, we
consider the  `gauge slice' $\widetilde S$  of
a partial gauge fixing in $\widetilde P'$:
\be
\widetilde S:= \{\, (A,E,\zeta)\,\vert \, \widetilde\Psi=0,\,\,
A(x) = \chi \in \check \cT^\gamma\,\}.
 \ee
Clearly the momentum map constraint $\widetilde \Psi=0$ requires
that the function $E(x)$ takes the form
\be
E(x) = e^{- (x - x_k) \ad_\chi } E_k^+
\quad\hbox{for}\quad
x_k < x < x_{k+1},
\label{Y17}\ee
where the only objects so far undetermined are the constants
\be
E_k^+\in \cG.
\label{fininf}\ee
On account of the quasi-periodicity, we have
$E_{k+N}^+ = \tau'(E_k^+)$ and we also define
\be
E_{k}^- = e^{- (x_{k} - x_{k-1}) \ad_\chi } E_{k-1}^+,
\qquad \forall k\in \bZ.
\ee
Since the function $E(x)$ has the jump $(E_k^+ - E_k^-)$ at $x_k$, we find
that when applied to the elements of $\widetilde S$
the constraint $\widetilde \Psi=0$ translates into
the following equations:
\be
\zeta_k + (E_k^+ - E_k^-) = 0,
\quad k=1\ldots, N.
\label{jumpcond}\ee
Now, our key observation is that the correspondence
 \be
q = \chi,\quad
\frac{1}{\lambda_k}=  x_{k} - x_{k-1},
\quad
\lambda_k \xi_k = \zeta_k,
\quad
\lambda_k J_k = E_{k-1}^+, \qquad k=1,\ldots, N,
\label{corresp}\ee
allows to reformulate (\ref{jumpcond}) as the following system of equations:
\be
\xi_k + \frac{\lambda_{k+1}}{\lambda_k} J_{k+1}
- e^{-\ad_{q/\lambda_k}} ( J_k)=0,
\qquad
k=1\ldots, N,
\ee
where  $J_{N+1}= \tau'(J_1)$ and $\lambda_{N+1}= \lambda_1$.
The crux  is that this  system of equations is nothing but
the componentwise form of the relation
(\ref{G13}), which represents the momentum map constraint $\Psi=0$
applied on the gauge slice $S$ in the finite-dimensional symplectic reduction
of Section 3.1.
Since the residual gauge transformations are in both cases given by the action
of the group $K$ (\ref{K}),
the relation (\ref{corresp}) induces a \emph{one-to-one correspondence} between the
reduced phase spaces coming from the infinite-dimensional and finite-dimensional
reductions:
\be
\widetilde P'_\red =\widetilde P'_{\widetilde \Psi=0}/ \widetilde G
 \simeq \widetilde S/K \simeq S/K \simeq P'_{\Psi=0}/G \simeq P'_\red.
\ee
It can be readily checked that \emph{the reduced symplectic structures and
also the reduced free Hamiltonians are converted into each other
upon this correspondence}.
The latter statement follows immediately since on $\widetilde S$ we have
\be
\int_0^1 \langle E(x), E(x)\rangle dx = \sum_{k=1}^{N} (x_{k} - x_{k-1})
\langle E_{k-1}^+,
E_{k-1}^+\rangle = \sum_{k=1}^N \lambda_k \langle J_k, J_k\rangle
= \langle \vJ, \vJ \rangle_{\vlambda}.
\ee
As for the symplectic form, let us first notice that the mapping
$\lambda_k \xi_k \leftrightarrow \zeta_k$
converts the (unreduced)
Poisson brackets in (\ref{G4}) into those in (\ref{4.6}).
In other words, $(\cO_k, \omega_k)$ and $(\cO'_k, \omega'_k)$
represent the same coadjoint orbit of $G$ in $\cG^*$,  identified with
$\cG$ via the different scalar products $\lambda_k \langle\ ,\ \rangle$ and
$\langle\ ,\ \rangle$.
Consequently, it is enough
to focus on the 1-form $\widetilde \theta$ (\ref{tildetheta}).
Regarding this, we have
\be
\widetilde \theta\vert_{\widetilde S} = \int_{0}^1 \langle E(x), d q \rangle dx
=\sum_{k=1}^{N} (x_{k}- x_{k-1}) \langle E_{k-1}^+, dq \rangle
= \langle \vJ, d \vq\rangle_{\vlambda}
= \langle \vJ_\cQ, d \vq \rangle_{\vlambda}
= \langle p, dq \rangle.
\ee
The second equality holds since only the $\cT$-part of $E(x)$ contributes, which is
constant on each interval $(x_{k-1}, x_k)$. We have taken into account the correspondence
(\ref{corresp}) together with equations (\ref{G15}), (\ref{G17}) and (\ref{T13}).
We conclude that the restriction of $\Omega$ on $S$ is mapped into
the restriction of $\widetilde \Omega$ on $\widetilde S$, which implies that
the respective reduced symplectic structures are converted into each other, as claimed.

The infinite-dimensional phase space that we started
with emerges from Yang--Mills theory in (1+1)-dimensions with
quasi-periodic boundary condition
and external non-dynamical charges located at the points $x_k$. Indeed, $A= A_1$ is
the spatial component of the Yang--Mills potential $A_\mu$ ($\mu=0,1$) in the gauge where
$A_0$ has been
set to zero, and $E$ is its canonical conjugate (the `color-electric'
field). Because of this interesting physical context \cite{GN-YM, BL1, Nekr, La}  it would be desirable to place the
above sketched derivation on completely rigorous mathematical ground.
We plan to deal with this issue, together with the isomorphism of the full reduced
phase spaces $P_{\Psi=0}/G$ and $\widetilde{P}_{\widetilde{\Psi}=0}/\widetilde{G}$,
elsewhere.

To finish, note from (\ref{tildeact}) that $A(x)$  transforms in the same way
as (at fixed level) the elements of the dual of a centrally extended twisted loop
algebra based on $\cG$.
In this section we followed the current algebraic derivation
of the Sutherland system as presented  in the book
\cite{Khes}, but (unlike there) we did not assume that $A(x)$ varies
in the \emph{smooth} dual.  Although such framework may appear
advantageous for the symplectic reduction itself, it
is problematic
since the flow of the `free Hamiltonian' $\widetilde H$ (\ref{tildeHam})
starting at $\widetilde S$ at $t=0$
leaves
the space of smooth fields $A$, because $E$ must have jumps at the points $x_k$.
In fact, our assumptions on $A$, $E$ and $g$ were designed to ensure that
the flow of the Hamiltonian $\widetilde H$,
\be
(A(x,t), E(x,t), \zeta_1(t),\ldots, \zeta_N(t)) = (A(x,0) + t E(x,0) , E(x,0),\zeta_1(0),\ldots, \zeta_N(0)),
\ee
stays in the unreduced phase space $\widetilde{P}$ on which $\widetilde{H}$ is well-defined and
enjoys $\widetilde{G}$-symmetry.

\section{Conclusion}
\setcounter{equation}{0}

In this paper we presented two derivations of novel spin Sutherland systems that
in special cases were studied earlier \cite{BL1,BL2,Pol}, and described the mechanism
whereby the two derivations always yield the same result.
The main virtues
of the infinite-dimensional derivation  are its connection
to Yang--Mills theory and that it can be extended for obtaining elliptic
generalizations,  similarly to the derivation
of the standard elliptic Calogero system from the current algebra on the torus
 \cite{GN-ell,Nekr,La}.
The finite-dimensional derivation has different advantages.
For example, it  permits the construction of  classical solutions
by a purely algebraic projection algorithm.  Its main advantage  is that, as we
outline next, the
corresponding quantum-Hamiltonian reduction is also in the range
of well-understood group-theoretic methods.

The quantum mechanical analogue of the unreduced classical system $(P,\Omega,H)$
is defined by the Hamilton operator
\be
\hat H = - \frac{1}{2} \Delta_{\vG} \otimes \mathrm{id}_{V_{\vnu}},
\ee
 where $\Delta_{\vG}$ is the Laplace--Beltrami operator associated with
 the Riemannian metric on $\vG$ that corresponds to
the invariant scalar product $\langle\ ,\ \rangle_{\vlambda}$.
Now this operator  acts on (its usual domain in)
the  Hilbert space
$L^2(\vG,  d\mu_{\vG}) \otimes V_{\vnu}$.
The measure $d\mu_{\vG}$ comes from the metric and the vector space
$V_{\vnu}$
carries  a highest weight representation $\rho_{\vnu}$ of the  direct
product group $\vG$. This representation is the `exterior tensor product'
of representations of the
$N$-factors,  which are quantum counterparts of the constituent orbits
$\cO_k$ of $\vcO$ (\ref{G1}), i.e.,
\be
V_\vnu=
V_{\nu_1}  \boxtimes \cdots \boxtimes V_{\nu_N}.
\label{D1}\ee

Any group element  $\veta\in \vG$ is represented by a unitary operator $U_\veta$ on the
Hilbert space, operating  on  a $V_\vnu$-valued function $F$ according to
\be
F \mapsto  U_\veta F = \rho_\vnu(\veta) \circ F \circ {\Act}^\Gamma_{{\veta}^{-1}}.
\label{D2}\ee
The reduced Hilbert space is provided by the subspace of $\vG$-singlets,
\be
\bigl(L^2(\vG,  d\mu_{\vG}) \otimes V_{\vnu}\bigr)^{\vG}.
\label{D3}\ee
This is mapped to itself by $\hat H$, and the corresponding restriction defines
the reduced quantum Hamiltonian.
By arguments similar to those in \cite{FPJPA}, it is not difficult to show
that the reduced quantum Hamiltonian is unitarily equivalent to the operator
$\hat H_\red$ given in (\ref{D5}) below that acts on (a suitable dense domain inside)
the Hilbert space
\be
L^2(\check \cT^\gamma, d\mu_{\check\cT^\gamma}) \otimes V_\vnu^{K},
\label{D4}\ee
where $V_\vnu^K\subset V_\vnu$  is formed by the vectors fixed by the subgroup
$K $ (\ref{K}) and the measure on $\check\cT^\gamma$ is defined by
the Euclidean scalar product $\langle\ ,\ \rangle$.
To describe the operator $\hat H_\red$, consider dual bases $T_a$ and $T^b$ of
$\K^\perp$  ($\langle T_a, T^b \rangle_\vlambda = \delta_a^b$) and
introduce $\cI^{a,b}(q) = \langle T^a, \cI(q) T^b \rangle_{\vlambda}$ using
$\cI(q)$ in (\ref{cI}).
Then one obtains
\be
\hat H_\red = - \frac{1}{2} \sum_i \frac{\partial^2}{\partial q_i^2}
+ \frac{1}{2}
\sum_{a,b} \cI^{a,b}(q) \rho_{\vnu}(T_a) \rho_\vnu(T_b) + \mathcal{C}_{G, N}^{\gamma,\vlambda},
\label{D5}\ee
where the $q_i$ are coordinates on $\cT^\gamma$ with respect to an orthonormal basis
and $\mathcal{C}_{G, N}^{\gamma,\vlambda}$ is a constant.
Observe that apart from this constant (which
arises from a similarity transformation \cite{FPJPA}),  the outcome of the quantum Hamiltonian reduction
is obtained from the classical Hamiltonian (\ref{G23})
via the naive quantization of the kinetic energy
and the replacement of
$\langle T_a, \vxi \rangle_{\vlambda}$  by  $\rho_\vnu(T_a)$
together with  restriction to the $K$-invariant subspace of $V_\vnu^K$
annihilated by the quantum analogues of the classical constraints enforcing
$\vxi_\K=0$.
The constant was calculated\footnote{A notational difference is that in \cite{FPNucl,FPJPA}
the normalized Killing form $\kappa$ was denoted by $\langle\ ,\ \rangle$.}
in \cite{FPJPA} for $N=1$ and any $\gamma$, and
now we also calculated it for the systems detailed in Section 3.2, i.e., for
$\gamma = \mathrm{id}_G$ and arbitrary $N$.  It turned out that
the constant is actually independent of $N$ and of
$\vlambda$ and it reads
$\mathcal{C}_{G, N}^{\mathrm{id},\vlambda}= - \frac{1}{2} \kappa(\delta, \delta)$
with the
`Weyl vector' $\delta = \frac{1}{2}\sum_{\alpha \in \Phi_+} \alpha$.

Finally, let us sketch how representation theory can be used
to determine, in principle, the spectrum of the reduced
Hamiltonian $\hat H_\red$. The clue is that the spectrum of $\Delta_{\vG}$ is
known from the Peter--Weyl theorem.
The theorem says that the representation of $\vG \times \vG$ on $L^2(\vG, d\mu_{\vG})$
that comes from left- and right-multiplications decomposes as
\be
L^2(\vG,d\mu_{\vG}) =\oplus_{\vLa} \left(V_{\vstarLa} \boxtimes V_\vLa\right),
\label{D6}\ee
where all `components' $\Lambda_k$ run over the dominant integral weights of $G$
and $\vstarLa$ is composed from the highest weights $\Lambda_k^*$, denoting the highest weight
of the contragredient of the representation $V_{\Lambda_k}$.
The Laplace--Beltrami operator is constant on each subspace
$V_{\vstarLa} \boxtimes V_\vLa$,
taking the value
\be
C_2^{\vlambda}(\vLa) =-\sum_{k=1}^N \lambda_k^{-1} \kappa( \Lambda_k + 2\delta,
\Lambda_k).
\label{D7}\ee
It is easy to see from this that the space of singlets
in $L^2(\vG,d\mu_{\vG})\otimes V_{\vnu}$ with respect to the representation (\ref{D2})
of $\vG$ can be presented as an infinite orthogonal direct sum of the finite-dimensional spaces
\be
(V_{\Lambda_2^*} \otimes V_{\Lambda_1} \otimes V_{\nu_1})^G
\boxtimes \cdots \boxtimes (V_{\Lambda_N^*} \otimes V_{\Lambda_{N-1}} \otimes V_{\nu_{N-1}})^G
\boxtimes (V_{\Lambda_1^*\circ \gamma'} \otimes V_{\Lambda_N} \otimes V_{\nu_N})^G.
\label{D8}\ee
The sum is over those vectors $\vLa$ for which the spaces
of the $G$-singlets in all the $N$ 3-fold tensor products in (\ref{D8}) are non-trivial.
The composition of $\Lambda_1^*$ with $\gamma'$ makes sense since $\gamma'$ acts
on the Cartan subalgebra on which the weights are defined.
The non-trivial finite dimensional spaces of the form (\ref{D8}) are
eigensubspaces
of the original Hamiltonian  $\hat H$, with eigenvalue
 $- \frac{1}{2} C_2^{\vlambda}(\vLa)$.
It is an archetypical Clebsch--Gordan problem to determine these spaces.
This infinite collection of finite-dimensional linear algebraic problems is equivalent
to the problem of  diagonalization for the generalized spin Sutherland
Hamiltonian $\hat H_\red$ (\ref{D5}).

\bigskip
\bigskip
\bigskip
\noindent{\bf Acknowledgements.}
The work of L.F. was supported in part by the Hungarian Scientific Research
Fund (OTKA) under the grant  K-111697.
He wishes to thank E.~Langmann and I.~Tsutsui
for useful discussions at an early stage of the work.
The work of B.G.P. was supported by the J\'anos Bolyai Research Scholarship of
the Hungarian Academy of Sciences. B.G.P was also supported by a Lend\"ulet Grant;
he is grateful to Z.~Bajnok for hospitality in the MTA Lend\"ulet Holographic QFT Group.

\renewcommand{\theequation}{\arabic{section}.\arabic{equation}}
\renewcommand{\theequation}{\arabic{section}.\arabic{equation}}
\renewcommand{\thesection}{\Alph{section}}
\setcounter{section}{0}

\section{Appendix on gauge transformations}
\setcounter{equation}{0}
\renewcommand{\theequation}{A.\arabic{equation}}

The argument presented below is adapted from the description
of the coadjoint orbits of the twisted affine Lie algebras \cite{W,MW},
which generalizes earlier results on the untwisted case.

Take a $\cG$-valued function $A$ on the real line $\bR$ that
satisfies the assumptions detailed in Section 4. In particular,
it is quasi-periodic in the sense that  $A(x+1) = \tau'(A(x))$, where $\tau'$ and $\tau$ are
 automorphisms of $\cG$ and $G$  induced from an automorphism of the Dynkin diagram.
 (The trivial automorphism is of course a special case.)
Then consider the following differential equation for a $G$-valued function $y_A$,
\be
y_A'(x) = y_A(x) A(x),
\ee
 where at the distinguished points $x_k$ (see Section 4) this is understood in the sense of one-sided limits.
 We further impose the initial condition
 \be
 y_A(0)= e\in G
 \ee
 and also require \emph{continuity of $y_A$ at all points $x_k$}.
 It is easily seen that there exists a unique `fundamental solution' $y_A$
 that meets these requirements.
 This  boils down to well-known existence and uniqueness statements
on each closed interval $[x_k, x_{k+1}]$, and
$y_A$ can be constructed by gluing the solutions
on each closed interval by continuity. The solution $y_A$ can be written
explicitly as an ordered product-integral (`Wilson line').
It follows that $y_A$ verifies the conditions used in our definition of the group
$\widetilde{G}$ in Section 4, \emph{except} for the quasi-periodicity condition.
Indeed, the quasi-periodicity of $A$ and the uniqueness of $y_A$
imply the relation
\be
y_A(x+1) = y_A(1) \tau(y_A(x)),
\qquad \forall x\in \bR.
\ee

Recall that there exists an element $w\in G$ and a unique
$\chi \in \mathrm{cl}({\check\cT^\gamma})$  for which
\be
e^\chi = C^\gamma_w(y_A(1)).
\ee
Here, $\mathrm{cl}({\check\cT^\gamma})$ is the closure of $\check \cT^\gamma$ introduced
at the beginning of Section 2. The automorphisms
$\gamma$ and $\tau$ are inverses of each other
(actually equal except possibly for $G=\mathrm{Spin}(8,\bR)$).
Picking $w$ and $\chi$, we now define the $G$-valued function $g_A$ on $\bR$ by
\be
g_A(x) = e^{-x \chi} \gamma(w) y_A(x).
\ee
It is trivial to check that $g_A$ satisfies the identity
\be
g_A(x) A(x) g^{-1}_A(x) - g_A'(x) g_A^{-1}(x) = \chi,
\qquad \forall x\in \bR,
\ee
and is  correctly quasi-periodic
\be
g_A(x+1) =\tau(g_A(x)).
\ee

In conclusion, we have shown that any function $A$ subject to our conditions
defining the  phase space $\widetilde{P}$ can be gauge transformed
into a constant $\chi$ from an arbitrarily chosen `twisted Weyl alcove' by
a gauge transformation defined by an element of the loop group
$\widetilde{G}$.

\end{document}